\documentclass{article}

\usepackage{arxiv}

\usepackage[english]{babel}
\usepackage[utf8]{inputenc}
\usepackage[T1]{fontenc}
\usepackage{graphicx}
\usepackage{calc}
\usepackage{hyperref}       
\usepackage{url}            
\usepackage{booktabs}       
\usepackage{amsfonts}       
\usepackage{nicefrac}       
\usepackage{microtype}      
\usepackage{natbib}
\usepackage{doi}
\usepackage{amsmath}
\usepackage{bm}
\usepackage{array}
\usepackage{float}
\usepackage{acronym}

\newcommand{\pd}[2]{\ensuremath{\frac{\partial #1}{\partial #2}}}
\newcommand{\fd}[2]{\ensuremath{\frac{\mathrm{d} #1}{\mathrm{d} #2}}}
\newcommand{\md}[2]{\ensuremath{\frac{\mathrm{D} #1}{\mathrm{D} #2}}}

\hypersetup{
    colorlinks = true,
    allcolors = black
}

\title{A Westervelt equation for acoustic wave propagation through weakly stratified, arbitrary Mach number atmospheres}

\author{ \href{https://orcid.org/0000-0003-3426-009X}{\includegraphics[scale=0.06]{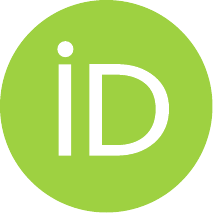}\hspace{1mm}Liam J.~Tope}\thanks{\href{mailto:L.J.Tope@soton.ac.uk}{L.J.Tope@soton.ac.uk}} \\
	Aerodynamics and Flight Mechanics Research Group \\
        University of Southampton \\
        Southampton SO17 1BJ, UK
	\And
	Jae Wook Kim \\
	Aerodynamics and Flight Mechanics Research Group \\
        University of Southampton \\
        Southampton SO17 1BJ, UK
        \AND
        Peter Spence \\
        AWE plc.\\
        Reading, RG7 4PR
}

\date{}

\renewcommand{\shorttitle}{A Westervelt equation for weakly stratified moving atmospheres}

\hypersetup{
pdftitle={A Westervelt equation for weakly stratified moving atmospheres},
pdfsubject={Nonlinear Acoustics},
pdfauthor={Liam J.~Tope, Jae Wook Kim, Peter Spence}
}

\begin{document}

 \newacro{ABL}{Atmospheric Boundary Layer}
 \newacro{ANAD}{Adaptive Nonlinear Artificial Dissipation}
 \newacro{CAA}{Computational Aeroacoustics}
 \newacro{CANARD}{Compressible Aerodynamics and Aeroacoustics Research code}
 \newacro{CFL}{Courant-Freidrichs-Lewy}
 \newacro{CTBT}{Comprehensive Nuclear-Test-Ban Treaty}
 \newacro{CTBTO}{Comprehensive Nuclear-Test-Ban Treaty Organisation}
 \newacro{DFT}{Discrete Fourier Transform}
 \newacro{DNS}{Direct Numerical Simulations}
 \newacro{FDTD}{Finite Difference Time Domain}
 \newacro{FFT}{Fast Fourier Transform}
 \newacro{HPC}{High Performance Computing}
 \newacro{HWM}{Horizontal Wind Model}
 \newacro{IMS}{International Monitoring System}
 \newacro{IVP}{Initial Value Problem}
 \newacro{MPI}{Message Passing Interface}
 \newacro{NRL-G2S}{Naval Research Laboratory Ground to Space}
 \newacro{NRLMSISE}{Naval Research Laboratory Mass Spectrometer and Incoherent Radar Model - Extended}
 \newacro{NWP}{Numerical Weather Prediction}
 \newacro{PDE}{Partial Differential Equation}
 \newacro{PE}{Parabolic Equation}
 \newacro{PML}{Perfectly Matched Layer}
 \newacro{PSD}{Power Spectral Density}
 \newacro{REPL}{Read–evaluate–print-loop}
 \newacro{SB}{Subdomain Boundary}
 \newacro{SSW}{Sudden Stratospheric Warming}
 \newacro{STP}{Standard Temperature and Pressure}
 \newacro{TL}{Transmission Loss}

\maketitle

\begin{abstract}
Nonlinear distortion of infrasonic waves through atmospheres up to thermospheric altitudes
govern large-range ground-level observations of explosive noise sources, causing
large differences between
the near and far field. Propagation modelling in this scenario to include realistic nonlinear
effects has thus far been limited to high-fidelity, numerically intensive Direct Numerical Simulations 
of the Navier-Stokes equations, or nonlinear parabolic equations with restrictions on the mean flow Mach
number. For the accurate modelling of nonlinear waveform synthesis through realistic atmospheric winds
up to thermospheric altitudes, this work presents nonlinear wave equation analysis which results in
in a Westervelt equation for weakly stratified, arbitrary Mach number atmospheres. This is intended to
be used as a benchmark for model development and numerical analyses such that alternative low-fidelity
numerical calculations to Direct Numerical Simulations can be sought.
\end{abstract}

\section{Introduction}

Within close vicinity to an explosion-source, acoustic pressure amplitudes can be several orders of magnitude 
larger than the ambient pressure level~\citep{kinney2013explosive}. Furthermore,
ambient atmospheric densities in the thermosphere are several orders of magnitude smaller than at ground level, 
resulting in large acoustic overpressure ratios.
Nonlinear propagation is therefore much more prevalent at such altitudes, 
resulting in N-wave formation~\citep{Hamilton1998} which heavily influences ground-level
observations~\citep{Sabatini2016,Marchiano2003} of the acoustic signal at large range.
Accurate models of the acoustic propagation within this environment is therefore dependent on the accurate
representation of nonlinear propagation effects.

\ac{DNS} of the Navier-Stokes equations have been commonly used in the infrasound propagation literature for
numerical predictions of remote infrasound observations. Such calculations permit general
propagation environments and a complete description of the acoustic field. Such an approach has demonstrated 
a capability for observing propagation effects
which linear, low-fidelity models do not replicate, including nonlinear waveform distortion 
with different source strengths~\citep{Marsden2014}, N-wave phase shifts through thermospheric caustics~\citep{Sabatini2016}, and
non-self-similar distortion of upward-propagating N-waves in the thermosphere~\citep{sabatini2019three}.
Despite this, the generality afforded by \ac{DNS} is granted by fine-scale numerical resolution of acoustic length-scales
over large distances, resulting in great computational expense. 

Alternative propagation models which include nonlinear effects have so far been based on nonlinear 
parabolic approximations, such as the KZK~equation~\citep{Kuznetsov1971}. Generalisations of the KZK~equation 
including flow motion to first-order~\citep{Averyanov2006} were used to study weak shock waves in turbulent 
flows\citep{Averiyanov2011}. Extensions beyond the narrow-angle limit of the standard parabolic approximation
have also been explored.
The generalized Lighthill-Westervelt equation has been used to develop numerical solutions
for shock wave diffraction through weakly heterogeneous media~\citep{Dagrau2011}; and
nonlinear wave equations~\citep{Coulouvrat2012} have been used to
extend this formulation to atmospheres with flow~\citep{Gallin2014}. These models permitted two-dimensional 
propagation through inviscid media. These were further extended to 3D with absorption and dispersions mechanisms 
associated with thermoviscosity and molecular relaxation~\citep{Luquet2019}.
For each of these models,
the numerical approach to solving the nonlinear parabolic equation was based on operator 
splitting~\citep{holden2010}.

The models discussed permit the inclusion of fine-scale flow 
heterogeneity as a result of an ordering scheme for the mean flow Mach number. For their proposed context of 
\ac{ABL} propagation, the restriction of low Mach number media is valid. However,
seasonal variation of zonal winds at stratospheric altitudes can yield wind-speed magnitudes which heavily
affect remote infrasound amplitude observations~\citep{mutschlecner1990}. In this work, a
wave equation for finite-amplitude atmospheric infrasound propagation which permits the use of strong
ambient winds is presented. It is intended that this provides a benchmark for further development of low-order
nonlinear propagation models pertinent to the long-range propagation of explosive infrasound sources. 
Such models which permit the efficient calculation of nonlinear atmospheric propagation
effects are highly sought for source determination from remote observations.

\section{Fluid Model}

Consider an inhomogeneous viscous and heat-conducting ideal gas, for which the
principle solution for the state variables is given by a sum of unperturbed and perturbed values. The unperturbed 
state variables correspond to the atmosphere in the absence of sound waves, whilst the perturbed fluid values are
associated with the acoustic field.

To find 
a solution, assumptions are made about spatio-temporal variations of the fluid. The atmospheric state is 
assumed to vary negligibly over acoustic timescales and horizontal spatial coordinates $x,y$.
The approximation of a purely stratified atmosphere has a reasonable
degree of accuracy for propagation over regional distances, on the order of a few hundred kilometres. 
Vertical ambient flows are also neglected, such that $\mathbf{u}_0 = \mathbf{u}_{0,H}$, where $H$ denotes the 
horizontal components. Turbulent fields or fine-scale atmospheric variations
induced by gravity waves can contribute to
non-zero vertical winds, however, these are of relatively small magnitude and so are assumed
to have a negligible effect on the infrasound ducting in the atmosphere.
The fluid state variables are therefore expressed as
\begin{equation}
    \begin{pmatrix}
        \rho \\ \mathbf{u} \\ p \\ T \\ s
    \end{pmatrix}
    = 
    \begin{pmatrix}
        \rho_0(z) + \rho^\prime(\mathbf{x},t) \\ \mathbf{u}_{0,H}(z) + \mathbf{u}^\prime(\mathbf{x},t) \\ p_0(z) + p^\prime(\mathbf{x},t)
        \\ T_0(z) + T^\prime(\mathbf{x},t) \\ s_0(z) + s^\prime(\mathbf{x},t)
    \end{pmatrix}~.
\end{equation}
Here the subscript $0$ denotes the ambient, or unperturbed state, and the $\prime$ superscript denotes the perturbation 
field. The field variables $T$ and $s$ denote the absolute temperature and specific entropy respectively.

\subsection{Governing Equations}
To describe fluid motions including the passage of sound through air, the physical conservation laws of a 
compressible, Newtonian fluid are considered. Namely, this includes the Continuity Equation for mass conservation
\begin{equation}\label{eq:nlwe-start-continuity}
    \md{\rho}{t} + \rho\bm{\nabla}\cdot\mathbf{u} = 0~,
\end{equation}
and the Navier-Stokes momentum conservation equation
\begin{equation}\label{eq:nlwe-start-momentum}
    \rho\md{\mathbf{u}}{t} + \bm{\nabla} p = \mu\nabla^2\mathbf{u} + (\mu_B + \frac{1}{3}\mu)\bm{\nabla}(\bm{\nabla}\cdot\mathbf{u})
        + \rho\mathbf{g}~,
\end{equation}
where the shear viscosity $\mu$ accounts for the diffusion of momentum between fluid parcels of different velocities, 
and the bulk
viscosity $\mu_b$ provides a low-frequency approximation of nonequilibrium deviations between the local and thermodynamic 
pressure. Here, the gravitational body force $g$ acts only in the $z$-direction towards the ground ($z=0$): $\mathbf{g}=(0,0,-9.81)$ m s$^{-2}$.
A complete account of all nonequilibrium departures requires molecular relaxation in Equation~\ref{eq:nlwe-start-momentum}. For this 
model it is assumed that relaxation times are much shorter than acoustic time scales, and are therefore
neglected. In both the continuity and momentum conservation equation, the material derivative $\md{}{t}$ has been used, which is 
given by $\md{}{t} = (\pd{}{t}+\mathbf{u}\cdot\bm{\nabla})$.

The form of the entropy equation is
\begin{equation}\label{eq:nlwe-start-entropy}
    \rho T\md{s}{t} = \kappa\nabla^2T  +\frac{\mu}{2}
        \sum_{i,j}\left(\pd{u_i}{x_j}+\pd{u_j}{x_i}-\frac{2}{3}\delta_{ij}\bm{\nabla}\cdot\mathbf{u}\right)^2
\end{equation}
for thermal conductivity $\kappa$, and Kronecker delta $\delta_{ij} = 1$ for $i=j$ and zero otherwise~\citep{pierce1989acoustics}.
Finally, the state equation is given here for an ideal gas,
\begin{equation}
    p = \rho RT
\end{equation}
where the thermodynamic pressure is a function of the density and entropy $p = p(\rho,s)$ instead of density and temperature
for convenience. This is a simplified model since the Equation of State for realistic atmospheres requires
relaxation mechanisms that significantly contribute to acoustic absorption at low altitudes. For lower frequencies this 
absorption is relatively small, and so for simplicity these are ignored.

\section{Ordering Scheme}\label{sec:nlwe-ordering}

In order to rank the importance of various terms in the governing equations, consider
a set of small quantities which define the relative scale of acoustic and ambient properties. 
Namely, the acoustic Mach number $\varepsilon = u/c_0$,
where $u$ is a typical acoustic velocity magnitude; a parameter measuring the importance
of viscous stresses $\eta = \mu\omega/\rho_0c_0^2$ and thermal conduction  $\eta / \Pr = \mu c_p / \kappa$, for characteristic 
angular frequency $\omega$ and Prandtl number $\Pr$; and a measure of the length scales of 
ambient medium inhomogeneities $\zeta = l_g / \lambda$, for characteristic wavelength $\lambda$ and the
characteristic length of stratified ambient fluid variations $l_g$.

Perturbations to the ambient field are all assumed to be of first order in $\varepsilon$, except for the entropy, which 
far from solid boundaries is $\mathcal{O}(\varepsilon^2)$~\citep{Hamilton1998}. Thermoviscous parameters $\mu$, $\mu_B$, 
and $\kappa$ are all $\mathcal{O}(\eta)$, and unlike the ambient state variables, they are assumed to have 
a constant value for all space and time. Finally, ambient fluid gradients are $\mathcal{O}(\zeta)$.
Terms in the governing equations which are proportional to one of these small quantities
is referred to as being linear. Terms which are proportional to the product of 
two quantities are second-order. Third-order or higher terms are considered negligible.

\section{Second Order Equations}

\subsection{Continuity Equation}
 Using the ambient fluid model, one can neglect the terms 
$\partial\rho_0/\partial t~=~\nabla\cdot\mathbf{u}_0~=~0$. The operator $\mathbf{u}_0\cdot\bm{\nabla}$ 
can also be simplified to only a sum of gradients along the horizontal directions
$\mathbf{u}_0\cdot\bm{\nabla} = \mathbf{u}_0\cdot\bm{\nabla}_H$. Rewriting the fluid state variables as the 
sum of the
unperturbed state and perturbations, expanding the material derivative operator, and exploiting these
properties, one has
\begin{equation}
    \pd{\rho^\prime}{t} + u_z^\prime\fd{\rho_0}{z} + \left(\mathbf{u}_0\cdot\bm{\nabla}_H\right)\rho^\prime + 
    \rho_0\bm{\nabla}\cdot\mathbf{u}^\prime + \left(\mathbf{u}^\prime\cdot\bm{\nabla}\right)\rho^\prime + 
    \rho^\prime\bm{\nabla}\cdot\mathbf{u}^\prime = 0.
\end{equation}
Using the definition $\mathcal{D}_H = \pd{}{t} + \mathbf{u}_0\cdot\bm{\nabla}_H$, and arranging $\mathcal{O}(\tilde{\varepsilon})$
terms on the left-hand-side (LHS) and $\mathcal{O}(\tilde{\varepsilon}^2)$ on the right-hand-side (RHS):
\begin{equation}\label{eq:nlwe-expanded-continuity}
    \mathcal{D}_H\rho^\prime + \rho_0\bm{\nabla}\cdot\mathbf{u}^\prime = - u_z^\prime\fd{\rho_0}{z}
    -\rho^\prime\bm{\nabla}\cdot\mathbf{u}^\prime - \left(\mathbf{u}^\prime\cdot\bm{\nabla}\right)\rho^\prime~.
\end{equation}

\subsection{Conservation of Momentum Equation}

Next consider the conservation of Momentum Equation~(\ref{eq:nlwe-start-momentum}). Expanded into the unperturbed field
and perturbations, one has
\begin{equation}
    \begin{split}
    \left(\rho_0+\rho^\prime\right)
    \left[\mathcal{D}_H\mathbf{u}^\prime
    + u_z^\prime\fd{\mathbf{u}_0}{z}
    + \left(\mathbf{u}^\prime\cdot\bm{\nabla}\right)\mathbf{u}^\prime \right]
    + \fd{p_0}{z}\mathbf{\hat{z}} + \nabla p^\prime = \mu\fd{^2\mathbf{u}_0}{z^2} + \mu\nabla^2\mathbf{u}^\prime \\ +
    \left(\mu_B+\frac{1}{3}\mu\right)\nabla\left(\bm{\nabla}\cdot\mathbf{u}^\prime\right) + (\rho_0 + \rho^\prime)\mathbf{g}~.
    \end{split}
\end{equation}
Using the mathematical identities
\begin{subequations}
    \begin{equation}
    \nabla\left(\bm{\nabla}\cdot\mathbf{u}\right) = \nabla^2\mathbf{u} + \bm{\nabla}\times\bm{\nabla}\times\mathbf{u}~,
\end{equation}
\begin{equation}
    \left(\mathbf{u}\cdot\bm{\nabla}\right)\mathbf{u} = \frac{1}{2}\nabla u^2 - \mathbf{u}\times\bm{\nabla}\times\mathbf{u}~,
\end{equation}
\end{subequations}
for $u^2 = \mathbf{u}\cdot\mathbf{u}$, recasts the equation as 
\begin{equation}
    \begin{split}
    \left(\rho_0+\rho^\prime\right)
    \left[\mathcal{D}_H\mathbf{u}^\prime
    + u_z^\prime\fd{\mathbf{u}_0}{z}
    + \frac{1}{2}\nabla {u^\prime}^2 - \mathbf{u}^\prime\times\bm{\nabla}\times\mathbf{u}^\prime \right]
    + \fd{p_0}{z}\mathbf{\hat{z}} + \nabla p^\prime = \mu\fd{^2\mathbf{u}_0}{z^2} + \mu\nabla^2\mathbf{u}^\prime \\ +
    \left(\mu_B+\frac{1}{3}\mu\right)\left(\nabla^2\mathbf{u}^\prime + \bm{\nabla}\times\bm{\nabla}\times\mathbf{u}^\prime\right) 
    + (\rho_0 + \rho^\prime)\mathbf{g}~.
    \end{split}
\end{equation}

The perturbations to the ambient state comprise three possible modes: acoustic, vorticity, and entropy modes. 
Weakly thermoviscous fluids have diffusive entropy and vorticity such that they decay 
exponentially from boundaries.
Linear theory approximates the curl of the perturbed velocity by 
$\bm{\nabla}\times\mathbf{u}^\prime \approx \bm{\nabla}\times\mathbf{u}^\prime_\mathrm{vor}$, and as a result 
these terms decay exponentially 
as the distance to the boundary increases. Following the treatment for a homogeneous, stationary
fluid~\citep{Hamilton1998}, the terms 
$\bm{\nabla}\times\bm{\nabla}\times\mathbf{u}^\prime$ and $\mathbf{u}^\prime\times\bm{\nabla}\times\mathbf{u}^\prime$ can be neglected. Using the gravitational acceleration $\mathbf{g} = [0, 0, -9.81]^T$ m 
s\textsuperscript{-1} and cancelling equal quantities associated with the hydrostatic equilibrium condition
\begin{equation}
    \fd{p_0}{z} = -\rho_0 g~,
\end{equation}
results in
\begin{equation}\label{eq:nlwe-momentum-final}
    \rho_0\mathcal{D}_H\mathbf{u}^\prime + \nabla p^\prime + \rho^\prime g = - \rho_0u_z^\prime\fd{\mathbf{u}_0}{z} 
    - \frac{\rho_0}{2}\nabla {u^\prime}^2
    - \rho^\prime\mathcal{D}_H\mathbf{u}^\prime + \mu\fd{^2\mathbf{u}_0}{z^2} 
    + \left(\mu_B+\frac{4}{3}\mu\right)\nabla^2\mathbf{u}^\prime~,
\end{equation}
when neglecting third-order or higher terms.

\subsection{Entropy Equation}

Expanding the entropy balance equation~(\ref{eq:nlwe-start-entropy}) into the unperturbed field and perturbations, 
a simplification is then made using $s^\prime\sim\mathcal{O}(\varepsilon^2)$ to 
cancel several fluid perturbation products. The final term in Equation~\ref{eq:nlwe-start-entropy} incorporating
viscosity and fluid flow products can also be neglected since the expanded form contains third-order
products of different combinations of $\varepsilon$, $\eta$, and $\zeta$. These simplifications with 
the assumed fluid model reduces to
\begin{equation}\label{eq:nlwe-entropy-final}
    \rho_0T_0\left(\mathcal{D}_Hs^\prime + u_z^\prime\fd{s_0}{z}\right) = \kappa\fd{^2T_0}{z^2} 
    + \kappa\nabla^2T^\prime~.
\end{equation}

\section{Equation of State}

The entropy equation can be refactored as a function of the density and pressure perturbations by using Taylor 
expansions of the Equation of State about the ambient state. To second-order, this is given by:
\begin{equation}\label{eq:nlwe-state-eqn-Taylor}
        p(\rho,s) = \left.p(\rho,s)\right|_0 + \left.\rho^\prime \pd{p}{\rho}\right|_0 
              + \left.\frac{{\rho^\prime}^2}{2}\pd{^2p}{\rho^2}\right|_0
              + \left.s^\prime\pd{p}{s}\right|_0 
\end{equation}
where the $0$ subscript denotes evaluation at ambient conditions. Using the definitions
\begin{subequations}
    \begin{equation}\label{eq:nlwe-ambient-sound-speed}
        \left.\pd{p}{\rho}\right|_0 = c_0^2
    \end{equation}
    \begin{equation}\label{eq:nlwe-b/a}
        \left.\pd{^2p}{\rho^2}\right|_0 = \frac{c_0^2}{\rho_0}\frac{B}{A}~,
    \end{equation}
\end{subequations}
where $c_0$ is the speed of sound in the unperturbed fluid and $B/A$ indicates the importance 
of the leading-order finite-amplitude correction to the small-signal sound speed~\citep{Hamilton1998},
then the Equation of State to second order is
\begin{equation}\label{eq:nlwe-state-eqn-prime-Taylor}
    p^\prime = c_0^2\rho^\prime + \frac{c_0^2}{\rho_0}\frac{B}{2A}{\rho^\prime}^2
              + \left.s^\prime\pd{p}{s}\right|_0~.
\end{equation}
Useful relations can be made between the results of the evaluation of the linear operator $\mathcal{D}_H$ since $c_0=c_0(z)$:
\begin{equation}\label{eq:nlwe-state-eqn-md}
    \mathcal{D}_Hp^\prime = c_0^2\mathcal{D}_H\rho^\prime 
                            + \frac{c_0^2}{\rho_0}\frac{B}{2A}\mathcal{D}_H{\rho^\prime}^2
                            + \left.\pd{p}{s}\right|_0\mathcal{D}_Hs^\prime~.
\end{equation}

These will prove useful at a later stage.

In the entropy equation~(\ref{eq:nlwe-entropy-final}), temperature perturbations $T^\prime$ are replaced with
the leading order term of the expansion
\begin{equation*}
    T^\prime = \rho^\prime\left.\pd{T}{\rho}\right|_0~,
\end{equation*}
as higher order terms are $\mathcal{O}(\tilde{\varepsilon}^3)$ in the entropy 
equation due the factor of $\kappa$. To evaluate $\partial T /\partial\rho$ and $\partial p/\partial s$ in 
Equation~\ref{eq:nlwe-state-eqn-prime-Taylor}, the Equation of State for an ideal gas is used
\begin{equation}
    \frac{p}{p_0} = \left(\frac{\rho}{\rho_0}\right)^\gamma \exp\left(\frac{s-s_0}{c_v}\right)~.
\end{equation}
The partial derivative in the expanded Equation of State is then given by
\begin{equation}
    \left.\pd{p}{s}\right|_0 = \frac{p_0}{c_v}~.
\end{equation}
The ideal gas law $p=\rho RT$ is then used to rewrite the Equation of State to
\begin{equation}
    T = \frac{p_0}{\rho_0^\gamma (c_p-c_v)}\rho^{\gamma-1}\exp\left(\frac{s-s_0}{c_v}\right)~,
\end{equation}
allowing one to define
\begin{equation}
    \left.\pd{T}{\rho}\right|_0 = (\gamma-1)\frac{T_0}{\rho_0}~.
\end{equation}
As a result, the temperature perturbations can be rewritten in terms of the density perturbations by
\begin{equation}
    T^\prime = \rho^\prime(\gamma-1)\frac{T_0}{\rho_0}~.
\end{equation}

Using this form for $T^\prime$ in the entropy equation~(\ref{eq:nlwe-entropy-final}) yields
\begin{equation}\label{eq:nlwe-entropy-tprime-sub}
    \rho_0T_0\left(\mathcal{D}_Hs^\prime + u_z^\prime\fd{s_0}{z}\right) = \kappa\fd{^2T_0}{z^2} 
    + (\gamma-1)\kappa\nabla^2\left(\rho^\prime\frac{T_0}{\rho_0}\right)~.
\end{equation}
The ambient entropy variations $\partial s_0/\partial z$ can also be substituted with ambient pressure 
and density variations using
\begin{equation*}
    \fd{p_0}{z} = \left.\pd{p}{\rho}\right|_0\fd{\rho_0}{z} + \left.\pd{p}{s}\right|_0\fd{s_0}{z}
\end{equation*}
\begin{equation*}
    \begin{split}
        \Rightarrow\hspace{2em}\left.\pd{p}{s}\right|_0\fd{s_0}{z} &= \fd{p_0}{z} - \left.\pd{p}{\rho}\right|_0\fd{\rho_0}{z} \\
        &= -\rho_0g - c_0^2\fd{\rho_0}{z}~.
    \end{split}
\end{equation*}
Rearranging equation~(\ref{eq:nlwe-entropy-final}) for $\mathcal{D}_Hs^\prime$, substituting into the
Equation of State~(\ref{eq:nlwe-state-eqn-md}), and using the ideal gas law $p_0/\rho_0T_0 ~=~ R ~=~ c_p-c_v$ and 
the definition of the heat capacity ratio 
$\gamma=c_p/c_v$, yields
\begin{equation}\label{eq:nlwe-state-final}
    \mathcal{D}_H p^\prime = c_0^2\mathcal{D}_H \rho^\prime + \frac{c_0^2}{\rho_0}\frac{B}{2A}\mathcal{D}_H {\rho^\prime}^2
                            + (\gamma-1)\kappa\fd{^2T_0}{z^2}
                            + \frac{c_0^2}{c_p\rho_0}(\gamma-1)\kappa\nabla^2\rho^\prime
                            + u_z^\prime\left(\rho_0g + c_0^2\fd{\rho_0}{z}\right)~,
\end{equation}
to second-order accuracy.

\section{Manipulation}

For an altitude-dependent
atmosphere with horizontal winds, the three equations up to second-order accuracy are 
the Continuity equation~(\ref{eq:nlwe-expanded-continuity}), the conservation of Momentum Equation~(\ref{eq:nlwe-momentum-final}),
and the state equation~(\ref{eq:nlwe-state-final}).
The corresponding first-order equations for linear, lossless acoustics are
\begin{subequations}\label{eq:nlwe-linear-lossless}
    \begin{equation}\label{eq:nlwe-linear-lossless-continuity}
        \mathcal{D}_H\rho^\prime + \rho_0\bm{\nabla}\cdot\mathbf{u}^\prime = 0
    \end{equation}
    \begin{equation}\label{eq:nlwe-linear-lossless-mom}
        \rho_0\mathcal{D}_H\mathbf{u}^\prime + \nabla p^\prime = 0
    \end{equation}
    \begin{equation}\label{eq:nlwe-linear-lossless-state}
        p^\prime = c_0^2 \rho^\prime~.
    \end{equation}
\end{subequations}
To derive an equation in terms of one fluid variable, the errors associated with the ordering scheme are exploited. The 
first-order terms incorporate an error of second-order, though when a first-order relation is 
substituted into a second-order term the error is of third-order. Only second-order or lower terms 
are considered in the present analysis, and therefore the errors introduced by this manipulation can 
are considered to be negligible. It is also 
acknowledged that gravity effects on acoustics fields are negligible at frequencies above $\sim$ 0.02 Hz, and thus 
can be ignored.

Using the first-order relation for $\bm{\nabla}\cdot\mathbf{u}^\prime$ from Equation~(\ref{eq:nlwe-linear-lossless-continuity}), the pressure-density relation in the linear state equation~(\ref{eq:nlwe-linear-lossless-state}), 
and the gradient of the acoustic pressure field from Equation~(\ref{eq:nlwe-linear-lossless-mom})
in the Continuity Equation~(\ref{eq:nlwe-expanded-continuity}), one finds
\begin{equation}
    \mathcal{D}_H\rho^\prime + \rho_0\bm{\nabla}\cdot\mathbf{u}^\prime = - u_z^\prime\fd{\rho_0}{z}
    +\frac{1}{2\rho_0c_0^4}\mathcal{D}_H{p^\prime}^2 + \frac{\rho_0}{2c_0^2}\mathcal{D}_H {u^\prime}^2~.
\end{equation}

Here, the Lagrangian energy density $\mathcal{L} = \mathcal{T} - \mathcal{V}$ for kinetic and potential energies $\mathcal{T}$
and $\mathcal{V}$ respectively can be used
\begin{equation*}
    \mathcal{L} = \frac{\rho_0{u^\prime}^2}{2} - \frac{{p^\prime}^2}{2\rho_0c_0^2}~,
\end{equation*}
to rewrite the Continuity Equation as
\begin{equation} \label{eq:nlwe-continuity-manip}
    \mathcal{D}_H\rho^\prime + \rho_0\bm{\nabla}\cdot\mathbf{u}^\prime = - u_z^\prime\fd{\rho_0}{z}
    + \frac{1}{c_0^2}\mathcal{D}_H\mathcal{L} + \frac{1}{\rho_0c_0^4}\mathcal{D}_H{p^\prime}^2~.
\end{equation}

Next the $\nabla^2\mathbf{u}^\prime$ term in the Momentum Equation is substituted for 
$\nabla\left(\bm{\nabla}\cdot\mathbf{u}^\prime\right)$ for only acoustic modes far from solid boundaries. 
Using Equation~(\ref{eq:nlwe-linear-lossless-mom}), Equation~(\ref{eq:nlwe-momentum-final}) becomes
\begin{equation}
    \rho_0\mathcal{D}_H\mathbf{u}^\prime + \nabla p^\prime = - \rho_0u_z^\prime\fd{\mathbf{u}_0}{z} 
    - \frac{\rho_0}{2}\nabla {u^\prime}^2 
    + \frac{\rho^\prime}{\rho_0}\nabla{p\prime}^2  + \mu\fd{^2\mathbf{u}_0}{z^2}
    + \left(\mu_B+\frac{4}{3}\mu\right)\nabla\left(\bm{\nabla}\cdot\mathbf{u}^\prime\right)~,
\end{equation}
Refactoring to include the Lagrangian density, and neglecting third-order terms introduced from ambient variations in 
$\nabla \mathcal{L}$, Equation~(\ref{eq:nlwe-momentum-final}) becomes
\begin{equation}\label{eq:nlwe-momentum-manip}
    \rho_0\mathcal{D}_H\mathbf{u}^\prime + \nabla p^\prime = -\nabla\mathcal{L} - \rho_0u_z^\prime\fd{\mathbf{u}_0}{z} 
    + \mu\fd{^2\mathbf{u}_0}{z^2} +\left(\mu_B+\frac{4}{3}\mu\right)\nabla\left(\bm{\nabla}\cdot\mathbf{u}^\prime\right)~.
\end{equation}

Finally, to rewrite the Equation of State, one can manipulate the linear relations via
$\mathcal{D}_H$(\ref{eq:nlwe-linear-lossless-continuity}) $- \bm{\nabla}\cdot$(\ref{eq:nlwe-linear-lossless-mom}).
Using the first-order relation 
\begin{equation*}
    \mathcal{D}_H\bm{\nabla}\cdot\mathbf{u}^\prime = \bm{\nabla}\cdot\mathcal{D}_H\mathbf{u}^\prime~,
\end{equation*}
this results in
\begin{equation}\label{eq:nlwe-linear-wave-eqn}
    \frac{1}{c_0^2}\mathcal{D}_H^2p^\prime - \nabla^2p^\prime = 0
\end{equation}
when retaining only first-order terms.

The Equation of State can be rewritten using the linear relation 
$p^\prime = c_0^2\rho^\prime$, and thus the manipulated linear equation~(\ref{eq:nlwe-linear-wave-eqn})
and linear Equation of State~(\ref{eq:nlwe-linear-lossless-state}) are used to derive
\begin{equation} \label{eq:nlwe-state-manip} 
    \frac{1}{c_0^2}\mathcal{D}_H p^\prime = \mathcal{D}_H \rho^\prime 
                                          + \frac{1}{\rho_0c_0^4}\frac{B}{2A}\mathcal{D}_H {p^\prime}^2
                                          + \frac{(\gamma-1)\kappa}{c_0^2}\fd{^2T_0}{z^2}
                                          - \frac{(\gamma-1)\kappa}{\rho_0c_0^4c_p}\mathcal{D}_H^2p^\prime
                                          + u_z^\prime\fd{\rho_0}{z}~.
\end{equation}

The manipulated equations~(\ref{eq:nlwe-continuity-manip}),~(\ref{eq:nlwe-momentum-manip}), 
and~(\ref{eq:nlwe-state-manip}) are now combined in the same manner as performed with the linear relations, by 
evaluating $\mathcal{D}_H$~(\ref{eq:nlwe-continuity-manip})~$- \nabla\cdot$~(\ref{eq:nlwe-momentum-manip}).
Using the second-order relation 
\begin{equation}
    \mathcal{D}_H\bm{\nabla}\cdot\mathbf{u}^\prime = \bm{\nabla}\cdot\mathcal{D}_H\mathbf{u}^\prime 
                                              - \fd{\mathbf{u}_0}{z}\cdot\nabla_Hu_z^\prime~,
\end{equation}
in this process yields
\begin{equation}\label{eq:nlwe-wave-eq-before-refinement}
    \mathcal{D}_H^2\rho^\prime - \nabla^2 p^\prime = \left(\frac{1}{c_0^2}\mathcal{D}_H^2 + \nabla^2\right)\mathcal{L}
                                            - 2\rho_0\fd{\mathbf{u}_0}{z}\cdot\bm{\nabla}_H u_z^\prime
                                            - (\mu_B + \frac{4}{3}\mu)\nabla^2(\bm{\nabla}\cdot\mathbf{u}^\prime)
                                            + \frac{1}{\rho_0c_0^4}\mathcal{D}_H^2 {p^\prime}^2~.
\end{equation}
Next, the material derivative $\mathcal{D}_H$~(\ref{eq:nlwe-state-manip})
is determined and the resulting 
value of $\mathcal{D}_H^2\rho^\prime$ is substituted into Equation~(\ref{eq:nlwe-wave-eq-before-refinement}). The linear relation $\bm{\nabla}\cdot\mathbf{u}^\prime = -\frac{1}{\rho_0c_0^2}\mathcal{D}_Hp^
\prime$ and
$\mathcal{D}_H\nabla^2p^\prime = \frac{1}{c_0^2}\mathcal{D}_H^3p^\prime$ from the manipulated linear 
equation~(\ref{eq:nlwe-linear-wave-eqn}) are further substituted into 
(\ref{eq:nlwe-wave-eq-before-refinement}). The $z$-component of the linear Momentum Equation $\mathcal{D}_Hu_z^
\prime = -\frac{1}{\rho_0}\pd{p^\prime}{z}$ is also substituted into the equation.
Lastly, defining the coefficient of nonlinearity $\beta=1+B/2A$ and 
diffusivity $ \delta = (\mu_B + \frac{4}{3}\mu)/\rho_0 + (\gamma-1)\kappa/(\rho_0 c_p)$, the wave equation 
is recast as
\begin{equation}\label{eq:AlmostTheWaveEquation}
    \frac{1}{c_0^2}\mathcal{D}_H^2p^\prime  - \nabla^2 p^\prime =
                            \left(\frac{1}{c_0^2}\mathcal{D}_H^2 + \nabla^2\right)\mathcal{L}
                          + \frac{\beta}{\rho_0c_0^4}\mathcal{D}_H^2 {p^\prime}^2
                          + \frac{\delta}{c_0^4}\mathcal{D}_H^3 p^\prime
                          - 2\rho_0\fd{\mathbf{u}_0}{z}\cdot\bm{\nabla}_H u_z^\prime -  \frac{1}{\rho_0}\fd{\rho_0}{z}\pd{p^\prime}{z}~.
\end{equation}

Equation~(\ref{eq:AlmostTheWaveEquation}) contains two variables describing the acoustic field, $p^\prime$ and $u_z^\prime$. 
Unless there is a known 
relationship between these field variables, this is not a suitable form for finding solutions. One can restrict the
wave equation to progressive, quasi-plane waves, such that
at distances greater than one wavelength from the source $\mathcal{L}$ is very small and can be neglected~\citep{Hamilton1998}. 
Furthermore, taking the material derivation $\mathcal{D}_H$ of the $z$-component of the linear conservation of Momentum Equation
allows one to recast $\mathcal{D}_H\nabla_Hu_z^\prime$. Therefore, taking the material derivative $\mathcal{D}_H$ of 
Equation~(\ref{eq:AlmostTheWaveEquation}) and substituting in the aforementioned linear relation, one finds
\begin{equation}\label{eq:nlwe_cross}
    \frac{1}{c_0^2}\mathcal{D}_H^3p^\prime  - \mathcal{D}_H\nabla^2 p^\prime =   
                          \frac{\beta}{\rho_0c_0^4}\mathcal{D}_H^3 {p^\prime}^2
                          + \frac{\delta}{c_0^4}\mathcal{D}_H^4 p^\prime -  \frac{1}{\rho_0}\fd{\rho_0}{z}\mathcal{D}_H\pd{p^\prime}{z}
                          + 2\fd{\mathbf{u}_0}{z}\cdot\nabla_H\pd{^2p^\prime}{z^2}~.
\end{equation}
This equation is cast in terms of only the acoustic pressure and is therefore amenable to calculating the acoustic field due 
to propagation through a weakly stratified moving atmosphere. Alternatively, to avoid several cross-derivative terms in favour
of simplicity, further approximations can be made. In particular, one can assume negligible wind shear $\fd{\mathbf{u}_0}{z} \ll \omega$
to obtain
\begin{equation}\label{eq:nlwe}
    \frac{1}{c_0^2}\mathcal{D}_H^2p^\prime  - \nabla^2 p^\prime =   
                          \frac{\beta}{\rho_0c_0^4}\mathcal{D}_H^2 {p^\prime}^2
                          + \frac{\delta}{c_0^4}\mathcal{D}_H^3 p^\prime -  \frac{1}{\rho_0}\fd{\rho_0}{z}\pd{p^\prime}{z}~.
\end{equation}

\section{Concluding remarks}

Equations~(\ref{eq:nlwe_cross}) and (\ref{eq:nlwe}) provide models for the propagation of infrasonic frequencies within weakly
stratified moving media within different wind shear regimes. This work has presented these analyses to serve as a benchmark
for further development of finite-amplitude atmospheric infrasound propagation, and reduce the reliance on \ac{DNS} for the 
accurate calculation of nonlinear effects on source-to-receiver waveform distortion. Furthermore, the increased efficiency of 
potential solutions to these equations indicates the feasibility of explosion source determination from remote observations, and
therefore is of importance for the characterisation of clandestine nuclear weapon test explosions.

\bibliographystyle{unsrtnat}
\bibliography{references}
\end{document}